\documentclass[aps,prl,twocolumn,footinbib,longbibliography]{revtex4-1}
\usepackage{amsmath,amssymb}
\usepackage{bm}
\usepackage{epsfig}
\usepackage{natbib}
\usepackage{hyperref}
\usepackage{color}
\usepackage{graphicx}

 \hypersetup{pdfstartview={FitH},pdfpagemode={UseNone},
            colorlinks,linkcolor=red, citecolor=blue, urlcolor=blue,
            bookmarks=true, bookmarksopen=true, pdfnewwindow=true}

\newcommand{\rmi}{{\rm i}}
\newcommand{\rmd}{{\rm d}}

\newcommand{\e}{{\rm e}}

\graphicspath{{./figs/}}

\begin{document}

\title{
Phonoritonic crystals with synthetic magnetic field for acoustic diode
}

\author{\firstname{A.~V.} \surname{Poshakinskiy}}
\email{poshakinskiy@mail.ioffe.ru}
\affiliation{Ioffe  Institute, St.~Petersburg 194021, Russia}
\author{\firstname{A.~N.} \surname{Poddubny}}
\affiliation{Ioffe  Institute, St.~Petersburg 194021, Russia}

\begin{abstract}

We develop a rigorous theoretical framework to describe light-sound interaction in the  laser-pumped periodic multiple-quantum-well structure accounting for the  hybrid phonon-polariton excitations, termed as phonoritons. We show that phonoritons exhibit the pumping-induced synthetic magnetic field in the  artificial ``coordinate-energy'' space, that makes transmission of left- and right- going  waves  different. The transmission nonreciprocity allows to use such {\it phonoritonic crystals} with realistic parameters as optically controlled nanoscale acoustic diodes. 

\end{abstract}

\maketitle

\paragraph{Introduction.}%
Reciprocity is the fundamental property of waves of different nature that means the invariance of the transmission coefficient under the interchange of the initial and final modes. Breaking of the reciprocity is required to build the devices, referred to as isolators or diodes, that allow wave transmission in one direction only~\cite{Maznev2013,Jalas2013}. Originating from the Onsager relations, reciprocity of linear transmission persists unless the time-inversion invariance is broken or the system is driven from the state of thermal equilibrium.

Violation of optical reciprocity can be achieved by different mechanisms, including nonlinearity~\cite{TrevioPalacios1996,Shi2015}, magneto-optical effect~\cite{Haldane2008,Shintaku1998,Bi2011,Inoue_book} and mode conversion~\cite{Li2014nc,Kim2015}. The proposed ways to achieve acoustic nonreciprocity are either based on nonlinearity~\cite{Li2004,Liang2009,Liang2010,Boechler2011,Popa2014,Chen2014,Liu2015,Zhang2015,Devaux2015,Gu2016}, that requires rather high signal powers;  mechanical rotation~\cite{Fleury2014}, temporal modulation of acoustic properties~\cite{Zanjani2014,Fleury2016} or introduction of a temperature gradient~\cite{Biwa2016}, that are difficult to realize at nanoscale.
We leave aside the mode-mismatched structures~\cite{Li2011}
that, while offering asymmetric transmission, are yet formally reciprocal and therefore cannot be used for the complete acoustic isolation~\cite{Maznev2013}.

Optomechanical resonators were shown to break reciprocity when driven by laser pump~\cite{Hafezi2012,Habraken2012,Schmidt2015,Peano2015,Ruesink2016,Shen2016}.
However, they require sophisticated fabrication while the problem of interfacing them with other acoustic devices remains unresolved. 
It was demonstrated recently that the photoelastic interaction in multiple-quantum-well (MQW) structure is greatly enhanced by exciton resonance~\cite{Jusserand2015}, boosting optomechanical effects in such systems~\cite{Poshakinskiy2016PRL}.    
We propose here to use laser-pumped MQWs as a nanoscale realization of an acoustic diode. %
Being easy-integrable with existing optoelectronic devices and semiconductor-based phonon lasers it promises various applications such as protecting the acoustic laser from backaction or even phononic computing~\cite{Sklan2015}.

Under laser pumping, hybrid phonon-polariton excitations, termed as {\it phonoritons}~\cite{Ivanov1982,Keldysh1986,Vygovskii1985,Hanke1999}, are formed. We show that the transport of phonoritons through the  1D MQW structure is equivalent to the quantum walk on a stripe of a 2D lattice in the virtual ``energy-coordinate'' space~\cite{Yuan2016}. Finite pump laser wave vector induces synthetic magnetic field on the 2D lattice, driving transmittance nonreciprocity. Proposed system can be viewed as a {\it phonoritonic crystal}, a periodic lattice
for interacting light and sound waves, as opposed to conventional photonic and phononic crystals. It can controllably amplify phonons transmitted in one direction while attenuating those transmitted in the opposite direction. Sound amplification and attenuation originate from the Brillouin scattering processes and can be related to the features of the phonoriton dispersion.

\begin{figure}[b!]
  \includegraphics[width=.99\columnwidth]{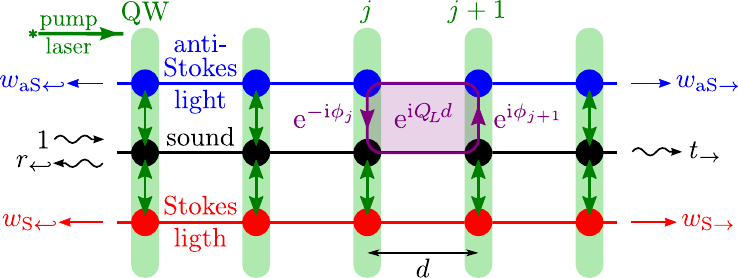}
\caption{Sound transmission through a multiple-quantum-well structure pumped by the laser from the left. 
Pumping-enhanced optomechanical interaction leads to sound conversion to anti-Stokes and Stokes light. The conversion is  equivalent to quantum walks on a $N \times 3$ lattice in a magnetic field with the flux $Q_Ld$ per plaquette, where $Q_L$ is the polariton wave vector at the pump frequency.
}\label{fig0}
\end{figure}

\paragraph{Exciton-mediated light--sound interaction.} 
The Hamiltonian describing interaction of excitons in a QW with light and longitudinal acoustic phonons reads  
\begin{align}\label{eq:H}
 H &= \omega_x b^\dag b+ \sum_q \omega_q^{\vphantom{\dag}} c_q^\dag c_q^{\vphantom{\dag}}  + \sum_k \Omega_k^{\vphantom{\dag}} a_k^\dag a_k^{\vphantom{\dag}} \\\nonumber
 &+ \sum_q \sqrt{\Gamma_0c}\, (c_q^\dag b + c_q^{\vphantom{\dag}} b^\dag) + \sum_k \frac{\rmi k \Xi_k}{\sqrt{2\rho\Omega_k S}} b^\dag b ( a_k^{\vphantom{\dag}} - a_k^\dag) \,.
\end{align}
The first line of the Hamiltonian~\eqref{eq:H} describes the bare excitons (bosonic annihilation operator $b$) with the energy $\omega_x$, free photons ($c_q$) and phonons ($a_k$) with the linear dispersion relations $\omega_q = c|q|$ and $\Omega_k = s|k|$, respectively, where $c$ and $s$ are the light and sound velocities in the medium, $q$ and $k$ are light and sound wave vectors,  $\hbar =1$. The first term in the second line describes the exciton-light interaction in the rotating-wave approximation. The interaction strength is determined by the exciton radiative decay rate  $\Gamma_0$~\cite{Ivchenko2005}. The last term of the Hamiltonian stands for the exciton-phonon interaction due to the deformation potential mechanism~\cite{Jusserand2015}. Here $\Xi_k$ is the deformation potential constant weighted with the exciton confinement wave function, $\rho$ is the mass density, and $S$ is the  sample area~\cite{Jusserand2013,Poddubny2014}.

When the QWs are pumped by a laser of the frequency $\omega_L$ close to exciton resonance, the coherent exciton polarization $b_L$ appears. 
Consider now sound transmission through such system.
The incident sound wave at the frequency $\Omega$ can be absorbed by this polarization giving rise to an optical wave at the frequency  $\omega_L+\Omega$, that we refer to as anti-Stokes. The amplitude of this conversion process is proportional to $b_L=|b_L|\e^{\rmi \phi}$ while the amplitude of conversion of the anti-Stokes light back into the sound is proportional to $b_L^*$. Alternatively, the incident sound wave can stimulate conversion of the laser-generated polarization to the Stokes light with the frequency $\omega_L-\Omega$; the amplitudes of the forth and back processes are proportional to $b_L^*$ and $b_L$, respectively.  The conversion occurs in all of $N$ QWs, so the initial wave transmission becomes equivalent to a quantum walk on a $N \times 3$ square lattice, see Fig.~\ref{fig0}. Importantly,
 conversion of the sound to Stokes light in the $(j+1)$th QW followed by back conversion in the neighboring $j$th QW yields a phase gain $\phi_{j+1}-\phi_j$, that is nonzero due to the finite pump laser wave vector. Such  process corresponds to the loop walk around the unit cell of the lattice; the gained phase can be viewed as  the flux of some effective magnetic field. This synthetic field induces the quantum Hall effect on the lattice, yielding nonreciprocal transport.

\paragraph{Transfer matrix approach.}
All the sound--light conversion processes that occur in laser-pumped QW can be described by $6\times 6$ scattering matrix relating the output and input particle operators  $(a_{k},a_{-k},c_{q_{{\rm aS}}}, c_{-q_{{\rm aS}}}, c_{q_{\rm S}}^\dag,c_{-q_{\rm S}}^\dag)$, where $k=\Omega/s$ and  $q_{{\rm aS}(S)} = (\omega_L \pm \Omega)/c$~\footnote{See Supplemental Material for the derivation of the scattering matrix elements by means of the Keldysh diagram technique.}. The fact that particle creation and annihilation operators are mixed by the scattering reflects the Bogolubov nature of the eigenstates of optomechanical system~\cite{Kippenberg2014}. 
For the study of multilayer system we use $6\times6$ transfer matrices, that relate phonon and photon operators on the right edges of the layer $(a_{k}^{\rm out},a_{-k}^{\rm in},c_{q_{{\rm aS}}}^{\rm out},c_{-q_{{\rm aS}}}^{\rm in},c_{q_{\rm S}}^{\rm out \,\dag},c_{-q_{\rm S}}^{\rm in \,\dag})$ to those on the left edge $(a_{k}^{\rm in},a_{-k}^{\rm out},c_{q_{{\rm aS}}}^{\rm in},c_{-q_{{\rm aS}}}^{\rm out},c_{q_{\rm S}}^{\rm in \,\dag},c_{-q_{\rm S}}^{\rm out \,\dag})$. The transfer matrix of a spacer has the diagonal $(\e^{\rmi kd}, \e^{-\rmi kd}, \e^{\rmi q_{\rm aS}d}, \e^{-\rmi q_{\rm aS}d}, \e^{-\rmi q_{\rm S}d}, \e^{\rmi q_{\rm S}d})$ and other elements are zero. The transfer matrix of a QW has a simpler form in the basis of acoustic and optical fields and their derivatives, $(a,a', c_{{\rm aS}}, c_{{\rm aS}}', c_{\rm S}^\dag, c_{\rm S}'^\dag)$, where $f = f_{q,\rm in(out)} + f_{-q,\rm out(in)}$ and $f' =\rmi[ f_{q,\rm in(out)} - f_{-q,\rm out(in)}]$ at the left (right) of the QW~\footnote{See Supplemental Material for the transfer matrix in the basis of traveling waves.}. It reads
\begin{align}\label{eq:TQW}
\hspace{-.2cm}\hat T
  =
\left(
\begin{array}{cccccc}
 1 & \frac{4 \gamma  \Delta }{(\Omega+\rmi\Gamma)^2-\Delta ^2} & \frac{-2 \e^{-\rmi \phi} \sqrt{\gamma  \Gamma_0 }}{\Omega+\rmi\Gamma+\Delta } & 0 & \frac{2 \e^{\rmi \phi} \sqrt{\gamma  \Gamma_0 }}{\Omega+\rmi\Gamma-\Delta } & 0 \\
 0 & 1 & 0 & 0 & 0 & 0 \\
 0 & 0 & 1 & 0 & 0 & 0 \\
 0 & \frac{2 \e^{\rmi \phi} \sqrt{\gamma  \Gamma_0 }}{\Omega+\rmi\Gamma+ \Delta} & \frac{2 \Gamma_0 }{ \Omega+\rmi\Gamma+ \Delta} & 1 & 0 & 0 \\
 0 & 0 & 0 & 0 & 1 & 0 \\
 0 & \frac{-2 \e^{-\rmi \phi} \sqrt{\gamma  \Gamma_0 }}{\Omega+\rmi\Gamma-\Delta } & 0 & 0 & \frac{-2 \Gamma_0 }{\Omega+\rmi\Gamma-\Delta } & 1 \\
\end{array}
\right)\hspace{-.1cm},
  \end{align}
where $\Delta=\omega_L-\omega_x$ is the laser detuning, $\gamma(\Omega)=|b_L|^2 |\Xi_k|^2 \Omega/(2\rho s^3 S)$ describes the strength of pumping-enhanced photoelastic interaction, $\Gamma$ is the nonradiative exciton decay rate, and $\e^{\rmi\phi}=b_L/|b_L|$ is the phase of the laser-generated exciton polarization. The structure of the matrix Eq.~\eqref{eq:TQW} reflects the fact that the QW exciton senses the optical fields $c_{{\rm aS}}$ and $c_{\rm S}^\dag$, and the derivative of the acoustic field (i.e., the deformation) $a'$, so they remain unchanged by the transfer matrix, while the components $a$, $c_{{\rm aS}}'$, and $c_{\rm S}'^\dag$ are discontinuous at the QW position.

We consider the sound transmission through the structure with $N$ QWs, see Fig.~\ref{fig0}. First,  using the conventional $2\times 2$ transfer matrix technique for light~\cite{Ivchenko2005} we calculate  the coherent exciton polarization $b_{L,i}$ that the pump laser creates in the $i$-th QW. Next, to calculate the sound transmission through the pumped MQW we use the $6\times 6$ transfer-matrix approach derived here: Transfer matrix for the whole structure $\hat T$ is obtained as a product of the transfer matrices of all the QWs and spacers. 
The coefficients of sound transmission from left to right (from right to left) $t_{\rightarrow(\leftarrow)}(\Omega)$  are readily given by formulae
\begin{align}\label{eq:tt}
 t_\rightarrow = \{[(\hat T^{-1})_{\rightarrow}]^{-1}\}_{11}\,, \quad t_\leftarrow = [(\hat T_{\leftarrow})^{-1}]_{11}\,, 
\end{align}
where $(\hat A)_{\rightarrow(\leftarrow)}$ denotes the $3\times3$ matrix formed by the matrix elements $\hat A$ in the basis of traveling waves with 
odd (even) row and column indices, i.e.,  corresponding to the modes with positive (negative) wave vector~\footnote{See Supplemental Materials for more details on the derivation of these formulae}.

\begin{figure}[t]
  \includegraphics[width=.99\columnwidth]{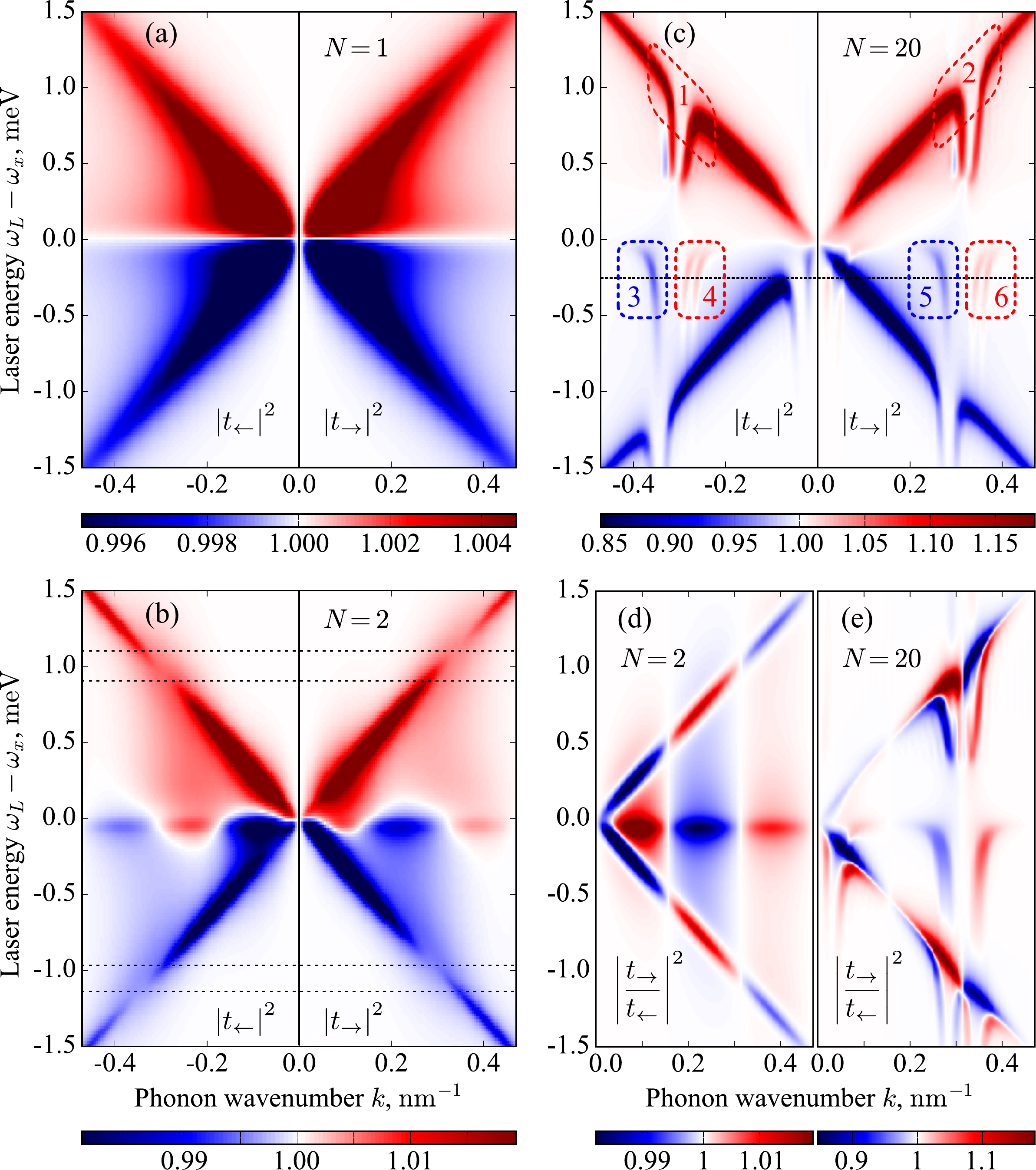}
\caption{(a)--(c) Dependence of the phonon transmission coefficient on the phonon wavenumber and pump laser frequency for structures with period $d=20$\,nm and different numbers of QWs $N$. Positive (negative) $k$ corresponds to left-to-right (right-to-left) transmission processes. (e) and (d) The ratio of the left-to-right and right-to-left transmittances. Calculation is made for $\Gamma = 50\,\mu$eV, $\Gamma_0=100\,\mu$eV, $\omega_x = 1.5\,$eV, and pump laser intensity 0.1\,mW per 1\,$\mu$m$^2$ cross section. Intrinsic phonon decay, acoustic and dielectric contrast are neglected. 
}\label{fig:tN}
\end{figure}

\paragraph{Nonreciprocal phonon transmission.}
Figures~\ref{fig:tN}(a)--(c) show the color maps of sound transmission coefficient through the  structures with different numbers of QWs $N$ as functions of the phonon wave vector $k$ and the laser detuning $\Delta$. Positive (negative) values of  $k$ represent the forward (backward) acoustic  transmittances $|t_{\rightarrow(\leftarrow)}(\Omega_k)|^2$. The transmission map for a single QW, Fig.~\ref{fig:tN}(a), has an X-like shape: The maximal sound amplification and attenuation is realized when the phonon energy matches the laser detuning $|\Delta|$~\cite{Poshakinskiy2016PRL}, in the analogy with the optomechanical heating and cooling~\cite{Kippenberg2014}. The map is symmetric with respect to the change of $k$ sign, meaning that the transmission through a single QW is reciprocal. This is because for a single QW there are no walk paths that enclose nonzero synthetic field flux in Fig.~\ref{fig0}.

Now we turn to the sound transmission through two QWs shown in Fig.~\ref{fig:tN}(b). The corresponding transmission map is not reciprocal: An asymmetric modulation arises on top of the  X-shaped feature. At small laser detuning the map is almost anti-reciprocal: when the right propagating phonon is amplified, the left-propagating phonon of the same frequency is attenuated and vice versa. At large laser detunings, the dips appear in the X-shaped feature. The dip position is different for $|t_\rightarrow|^2$ and $|t_\leftarrow|^2$, see the dashed lines in Fig.~\ref{fig:tN}(b). 
The nonreciprocal transmission pattern originates from the walk trajectories in Fig.~\ref{fig0} that involve conversion of sound to Stokes or anti-Stokes light in one QW followed by back conversion in the other QW. The counterclockwise (clockwise) loop paths enclose the synthetic field flux $\pm q_Ld$, where $q_L=\omega_L/c$. 
Summation of the amplitudes corresponding to 4 elementary trajectories enclosing nonzero flux gives~\footnote{See Supplemental Materials for the derivation of the amplitudes.}
\begin{align}\label{eq:dt}
 \delta|t_\rightarrow|^2 = 2{\rm Re}\sum_{\sigma=\pm} \frac{-\sigma\gamma\Gamma_0 
 }{(\sigma\Delta+\Omega+\rmi\Gamma)^2} \cos (\sigma q_Ld+ kd) .
\end{align}
Here the terms with $\sigma=+(-)$ correspond to the processes with anti-Stokes (Stokes) intermediate states that dominate at positive (negative) detunings, and $q_{\rm aS(S)} d\ll 1$ was supposed.
The asymmetry between the transmission for left- and right- incident phonons is driven by  the pump laser that, being always incident from the left, induces the synthetic magnetic field on the lattice Fig.~\ref{fig0}. In terms of the quantum walk, for any path contributing to $t_{\rightarrow}$ that encloses synthetic field flux $q_Ld$, the reciprocal path contributing to $t_{\leftarrow}$ will enclose the flux $-q_Ld$.  Therefore, the  reciprocal transmittance $\delta|t_{\leftarrow}|^2$ is obtained from Eq.~\eqref{eq:dt} by reversing the sign of $q_L$, that plays the role of magnetic field.
The ratio of the transmittances $|t_\rightarrow/t_\leftarrow|^2 \approx 1+\delta|t_\rightarrow|^2 - \delta|t_\leftarrow|^2$ is shown in Fig.~\ref{fig:tN}(d). In the relevant case of small $q_Ld\ll1$ it is proportional to $q_L d \sin kd$ and has extrema at $\omega_L = \omega_x$ (resonance for input light) and $\omega_L = \omega_x \pm \Omega$ (resonance for scattered light)~\cite{Poddubny2014}. 

\begin{figure}[t]
  \includegraphics[width=.99\columnwidth]{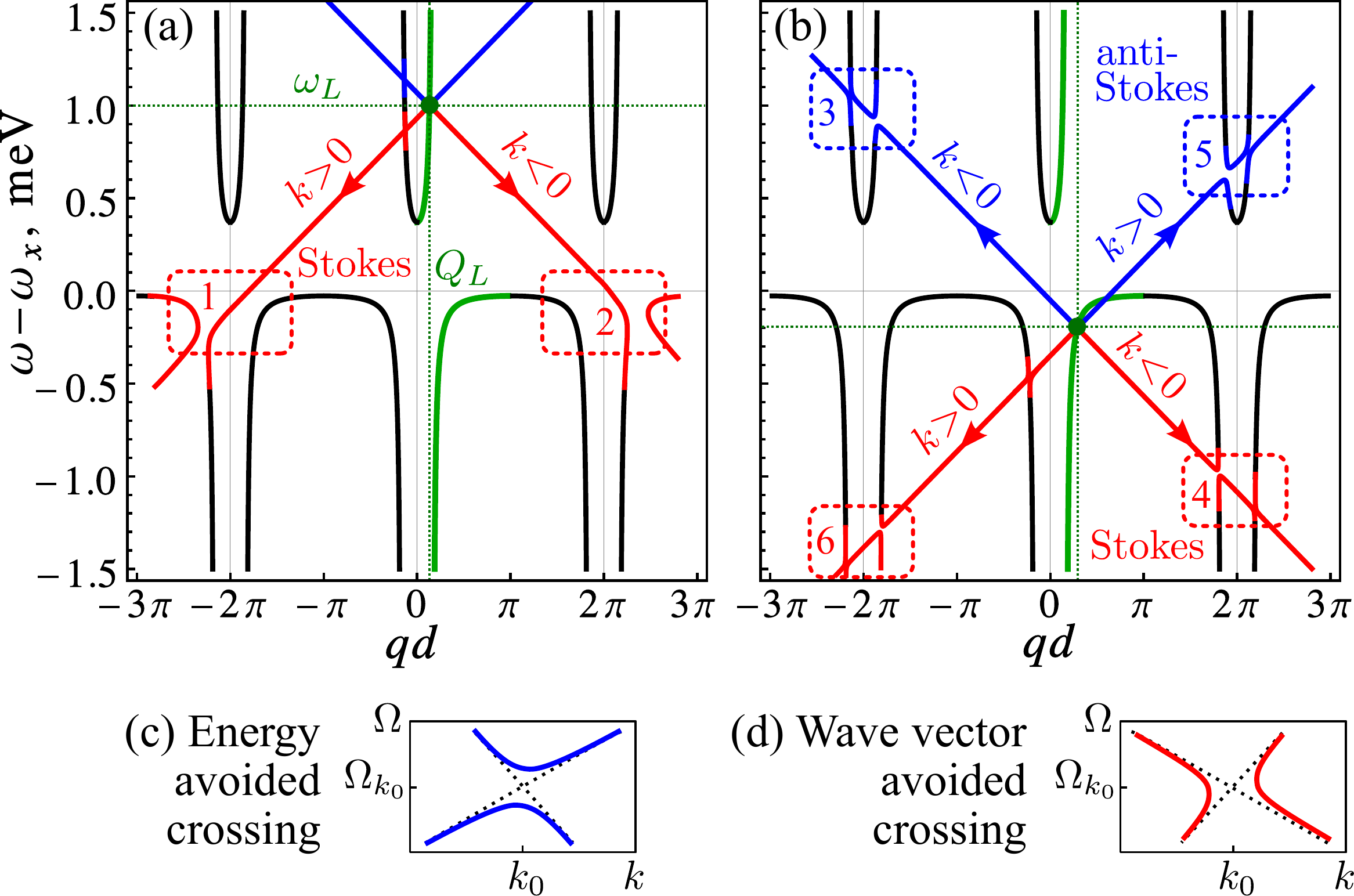}
\caption{Panels (a) and (b): Dispersion of polaritons (black curves),   emitted and absorbed phonons (red and blue lines) and 
phonoritons (avoided crossings at the intersections  1--6 of the polariton and phonon dispersions).  Panels (c) and (d) illustrate the two types of the avoided crossings in the dispersion, marked by blue and red color in panels (a)-(b), and yielding attenuation and amplification of transmitted phonons, respectively. The  areas 1--6 in panel (a)-(b) explain the corresponding patterns in the phonon transmission map Fig.~\ref{fig:tN}(c). The calculation is made for laser detunings $\Delta=1\,$meV and $\Delta=-0.2\,$meV [panels (a) and (b), respectively]; the other parameters are the same as for Fig.~\ref{fig:tN} except for except for $\Gamma=0$ and $\gamma(1\,\text{meV})=20\,\mu$eV. 
}
\label{fig:fig_j}
\end{figure}

It follows from Eq.~\eqref{eq:dt} that the dips in the X-like feature appear at $|k|=\pi/d\pm q_L$. With increase of the number $N$ of QWs the dips enlarge, leading to the complete suppression of amplification/attenuation in a certain interval, see regions 1 and 2 in Fig.~\ref{fig:tN}(c). 
We explain an appearance of this gap as well as the complex pattern of the transmittance in regions 3--6 by formation of the exciton-polaritons in long structures. These hybrid excitation are caused by interaction of excitons and light in MQW structures~\cite{Ivchenko2005}. The polariton dispersion in the infinite lossless MQW structure is shown by the black lines in Fig.~\ref{fig:fig_j}(a). It can be viewed as a result of an avoided crossing of the exciton dispersion, that is a horizontal line $\omega=\omega_x$, with the light dispersion, that is almost vertical, $\omega=c|q|$. The polariton dispersion is $2\pi/d$-periodic in the wave vector due to the translational symmetry of the infinite MQW structure.

The pump laser generates the coherent polariton population with the wave vector $Q_L = (1/d) \arccos [ \cos q_L d + (\Gamma_0/\Delta) \sin q_L d ]$~\cite{Ivchenko2005} that follows right-propagating polariton dispersion, see thick green curve. When a phonon propagates though such pumped structure it can either be absorbed in a process of anti-Stokes polariton scattering or stimulate emission of an additional phonon in the process of Stokes scattering.
Figure~\ref{fig:fig_j}(a) shows by blue and red lines the anti-Stokes and Stokes polariton scattering processes, respectively, for the case when laser energy is $\omega_L=\omega_x+1\,$meV.
Intersections of the blue and red lines, that represent the dispersions of the absorbed and emitted phonons, respectively, with the polariton dispersion (black line) determine all possible scattering processes~\cite{Poddubny2014}. These processes are then manifested in the transmittance map: transmitted phonons that are resonant with anti-Stokes scattering processes are attenuated while those resonant with Stokes process get amplified. Therefore, the pattern of the transmittance map Fig.~\ref{fig:tN}(c) mimics the  polariton dispersion.

In particular, the discussed above gaps in the transmission amplification, regions 1 and 2 in Fig.~\ref{fig:tN}(c), reflect the polariton gap, that suppresses the  Stokes scattering process at the corresponding frequency regions, marked in Fig.~\ref{fig:fig_j}(a). Similarly to the case of two QWs, the position of the gap for right- and left-propagating phonons is different due to nonzero laser-generated polariton wave vector $Q_L$. 
Figure~\ref{fig:fig_j}(b) sketches the Brillouin scattering processes for the laser energy $\omega_L=\omega_x-0.2\,$meV. Both Stokes processes (3 and 5) and anti-Stokes processes (4 and 6) are revealed in the acoustic transmittance map as attenuation and amplification, see the corresponding regions in Fig.~\ref{fig:tN}(c). The transmittance map is strongly nonreciprocal: Inside the frequency band of right-propagating phonon amplification 6 the left-propagating phonons are attenuated (region 3), and vice versa with regions 4 and~5.

 \paragraph{Phonoritons.} At the vicinity of the points in Fig.~\ref{fig:fig_j}(a), where the polariton and phonon dispersions intersect,  the strong photoelastic interaction modifies the eigenmodes of the system that become now phonoritons, a hybrid of phonon and polariton~\cite{Ivanov1982}, so
far considered only in bulk crystals. 
The dispersion of phonoritons in the infinite MQW structure without losses can be found from the equation $\Omega^2 = \Omega_k^2 + 2\Omega_k \Sigma_k$, where the phonon self-energy correction $\Sigma_k$ is readily expressed via the polariton Green's function,
\begin{align}
&\Sigma_k=\frac{\gamma s}{d}[G(Q_L+k,\omega_L+\Omega) + G(Q_L-k,\omega_L-\Omega)]\,,\nonumber\\\label{eq:phondisp}
&G(K,\omega) = [\omega-\omega_x-\Gamma_0 \sin qd/(\cos Kd-\cos qd)]^{-1}\hspace{-.097cm}.
\end{align}
Photoelastic interaction splits the phonoriton energy levels at the crossing points of phonon and polariton dispersions. In the vicinity of a crossing at a phonon wave vector $k_0$,  Eq.~\eqref{eq:phondisp} can be linearized yielding the phonoriton dispersion $(\Omega-\Omega_k)[\Omega-\Omega_{k_0}-v_g(k-k_0)]=\delta^2$. The value of energy level splitting $2\delta$ is determined by the equation
\begin{align}\label{eq:delta}
\delta^2 = \pm \frac{\gamma v_g s }{d^2} \frac{\sin qd}{\sin Qd} \frac{\Gamma_0}{(\omega-\omega_x)^2} \,,
\end{align}
where $+$ or $-$ sign should be chosen depending whether the crossing of polariton with anti-Stokes or Stokes phonon branches is considered, $v_g=(\rmd Q/\rmd\omega)^{-1}$ is the polariton group velocity and all the quantities should be taken at the frequency of the crossing. 

It follows from Eq.~\eqref{eq:delta} that for the crossings of polaritons with anti-Stokes phonons the splitting value $\delta$ is real, while for those with Stokes phonons it is imaginary.
This results in the two types of avoided level crossings~\cite{Wang1990} that we refer to as energy avoided crossings and wave vector avoided crossings, and mark by blue and red color, respectively. 
The energy avoided crossing ($\delta^2>0$) is the conventional one that corresponds to appearance of a gap for certain frequency range, see Fig.~\ref{fig:fig_j}(c). Propagation of the phonons with the frequencies inside the gap into the bulk of the structure is suppressed, therefore such avoided crossing leads to the attenuation of transmitted phonons. The wave vector avoided crossing ($\delta^2<0$) is the one when the gap appears for certain wave vector range, see Fig.~\ref{fig:fig_j}(d). The gap is constrained by two exceptional points; the states with the wave vectors inside the gap have complex energies. That means they can grow in time, therefore the infinite system is unstable. The finite structure that is not too long is yet stable, but the transmitted phonons are amplified~\cite{Poshakinskiy2016PRL}.

The phonoriton gaps shown in Fig.~\ref{fig:fig_j} correspond to Umklapp polariton scattering processes that involve phonons with large wave vectors $k\approx 2\pi/d$ and, therefore, stronger deformation potential~\cite{Poddubny2014}. Enhancement of the phonoriton effect due to the spectrum folding is the advantage of MQWs over bulk crystals.
Still, the value of splitting at avoided crossings, being proportional to the small parameter $\gamma$, for realistic structure parameters it is much smaller than the nonradiative exciton decay $\Gamma$. Therefore, even in the state-of-the-art structures the discussed above peculiarities of phonoriton dispersion are greatly broadened. Nevertheless, the attenuation and amplification of the transmitted sound remain in a lossy structure as an afterglow of the energy and wave vector avoided crossings, correspondingly, cf. the numbered areas in Figs.~\ref{fig:tN}(c) and~\ref{fig:fig_j}(a).

To summarize, the exciton-enhanced light-sound interaction leads to the formation of hybrid phonoriton modes in the laser-pumped periodic array of quantum wells. Propagation of such modes in the structure is equivalent to the quantum walk on a square lattice in external magnetic field and can be described by a $6\times 6$ transfer matrix technique. The strongly nonreciprocal spectra of phonoriton modes lead to nonreciprocal sound transmission. The proposed {\it phonoritonic crystal} can be used as an on-chip acoustic diode with optically controllable properties, that is easy integrable with existing optoelectronic devices.

\paragraph*{Acknowledgments.} 
The authors acknowledge fruitful discussions with  A. Fainstein and S.G. Tikhodeev.
This work was supported by
the RFBR and the Foundation ``Dynasty.'' A.N.P. and A.V.P.
also acknowledge support by the Russian President Grants No.
MK-8500.2016.2 and SP-2912.2016.5, respectively.

\onecolumngrid

\setcounter{equation}{0}
\setcounter{figure}{0}
\setcounter{table}{0}
\makeatletter
\renewcommand{\theequation}{S\arabic{equation}}
\renewcommand{\thefigure}{S\arabic{figure}}
\renewcommand{\thesection}{S\arabic{section}}

\vspace{.02\paperheight}
\newpage

\begin{center}
\textbf{\large Supplemental Material\\ for ``Phonoritonic crystals with synthetic magnetic field for acoustic diode''}
\end{center}

\subsection{Phonon-photon scattering matrix of a single QW}

 Green's function of a QW exciton dressed by the interaction with phonons reads~\cite{Poshakinskiy2016PRL}
 \begin{align}
\hat G^R (\omega)= \left[ \begin{array}{ll}
G^R(\omega)  & G^{\diamond}(\omega) \\
G^{\times}(\omega) & G^A(2\omega_L-\omega)
         \end{array} \right]
         = \frac{1}{\Delta^2-(\Omega + \rmi\Gamma_x)^2 - 2\Delta\Sigma}
  \left[ \begin{array}{cc}
\Delta-\Omega-\rmi\Gamma_x-\Sigma  & \Sigma\e^{2\rmi\phi}\\
\Sigma\e^{-2\rmi\phi} & \Delta+\Omega+\rmi\Gamma_x-\Sigma
         \end{array} \right] \,,
\end{align}
Here $\Omega = \omega-\omega_L$, $\Delta=\omega_L-\omega_x$ is the laser detuning, $\Gamma_x = \Gamma + \Gamma_0$ is the total exciton decay rate, $G^R$ and $G^A$ are the retarded and advanced exciton propagators, $G^{\diamond}$ and $G^{\times}$ are the Beliaev propagators corresponding to the creation of an exciton pair from the condensate of excitons generated by laser, and its annihilation, respectively~\cite{Keldysh1986}. We suppose that the real part of the exciton self-energy $\Sigma$ is already included in the exciton resonance frequency and leave only the imaginary part, $ \Sigma(\Omega) = - \rmi \gamma (\Omega)$.

The phonon reflection and transmission coefficients for a single QW were previously calculated in Ref.~\cite{Poshakinskiy2016PRL} and read
\begin{align}
&r(\Omega) = \frac{2\Delta\rmi\gamma }{\Delta^2 - (\Omega+\rmi\Gamma_x)^2+2\Delta\rmi\gamma }\,,\label{tk}\\
&t(\Omega) = 1-r(\Omega) \,.\label{rk}
\end{align}
The amplitudes of phonon conversion into an anti-Stokes photon and of its annihilation with a Stokes photon are readily calculated as,
\begin{align}\label{sa}
&w_{{\rm aS}}(\Omega) = \sqrt{\gamma\Gamma_0}(G^R \e^{\rmi\phi} + G^{\diamond} \e^{-\rmi\phi} )=\e^{\rmi\phi}\sqrt{\gamma\Gamma_0}\,\frac{\Delta-\Omega-\rmi\Gamma_x}{\Delta^2 - (\Omega+\rmi\Gamma_x)^2+2\Delta\rmi\gamma }\,,\\\label{ss}
&w_{{\rm S}}(\Omega) =  -\sqrt{\gamma\Gamma_0}(G^A \e^{-\rmi\phi} + G^{\times} \e^{\rmi\phi} )=-\e^{-\rmi\phi}\sqrt{\gamma\Gamma_0}\,\frac{\Delta+\Omega+\rmi\Gamma_x}{\Delta^2 - (\Omega+\rmi\Gamma_x)^2+2\Delta\rmi\gamma }\,.
\end{align}
The amplitudes Eq.~\eqref{sa}--\eqref{ss} are written for the case of right-propagating phonon ($k>0$); for $k<0$ they are $-w_{{\rm aS}}$ and $-w_{{\rm S}}$. The reflection and transmission coefficients for anti-Stokes and Stokes light read
\begin{align}
&r_{{\rm aS}}(\Omega) = -\rmi\Gamma_0 G^R = -\rmi\Gamma_0\,\frac{\Delta-\Omega-\rmi\Gamma_x+\rmi\gamma}{\Delta^2-(\Omega + \rmi\Gamma_x)^2 + 2\Delta\rmi\gamma} \,,\\
&r_{{\rm S}}^*(\Omega) = \rmi\Gamma_0 G^A = \rmi\Gamma_0\,\frac{\Delta+\Omega+\rmi\Gamma_x+\rmi\gamma}{\Delta^2-(\Omega + \rmi\Gamma_x)^2 + 2\Delta\rmi\gamma} \,,\\
&t_{{\rm aS(S)}}(\Omega) = 1+r_{{\rm aS(S)}}(\Omega) \,.\label{rk}
\end{align}
The amplitudes of the processes of anti-Stokes  conversion into a phonon and of creation of a phonon and a Stokes photon are  $-\bar w_{{\rm aS}}$ and $-\bar w_{{\rm S}}$ where the bar indicates inversion of the sign of $\phi$. The amplitude of annihilation of an anti-Stokes and a Stokes photon read
\begin{align}
w_{{\rm S}aS}(\Omega) = \rmi\Gamma_0 G^\times = \frac{\gamma \Gamma_0 \e^{-2\rmi\phi}}{\Delta^2 - (\Omega+\rmi\Gamma_x)^2+2\Delta\rmi\gamma } \,,
\end{align}
while the amplitude of creation of such pair is $-\bar w_{{\rm S}aS}$.

Note that in the absence of nonradiative losses, $\Gamma = 0$, the coefficients fulfill the conservation laws, 
\begin{align}
& |r|^2 + |t|^2 + 2 |w_{{\rm aS}}|^2 - 2 |w_{\rm S}|^2 = 1 \,,\\
& |r_{{\rm aS}}|^2 + |t_{{\rm aS}}|^2 + 2 |w_{{\rm aS}}|^2 - 2 |w_{{\rm S}aS}|^2 = 1 \,,\\
& |r_{{\rm S}}|^2 + |t_{{\rm S}}|^2 - 2 |w_{{\rm S}}|^2 - 2 |w_{{\rm S}aS}|^2 = 1 \,,
\end{align}
that follow from conservation of the quantity $N_{\rm phonon} + N_{\rm anti\text{-}Stokes} - N_{\rm Stokes}$.
The factor 2 reflects two possible propagation directions of the final particles.

All the above coefficients form the scattering matrix $\hat S$, relating input and output operators:
\begin{align}\label{eq:SQW}
 \left(\begin{array}{l}
 a_{k} \\ a_{-k} \\ c_{q_{{\rm aS}}} \\ c_{-q_{{\rm aS}}} \\ c_{q_{{\rm S}}}^\dag \\ c_{-q_{{\rm S}}}^\dag
 \end{array}\right)^{\rm out} = \quad
 \left(\begin{array}{rrrrrr}
  t & r & -\bar w_{{\rm aS}} & -\bar w_{{\rm aS}} & \bar w_{{\rm S}} & \bar w_{{\rm S}} \\
  r & t & \bar w_{{\rm aS}} & \bar w_{{\rm aS}} & -\bar w_{{\rm S}} & -\bar w_{{\rm S}} \\
  w_{{\rm aS}} & -w_{{\rm aS}} & t_{{\rm aS}} & r_{{\rm aS}} & -\bar w_{{\rm S}aS} & -\bar w_{{\rm S}aS} \\
  w_{{\rm aS}} & -w_{{\rm aS}} & r_{{\rm aS}} & t_{{\rm aS}} & -\bar w_{{\rm S}aS} & -\bar w_{{\rm S}aS} \\
  w_{{\rm S}} & -w_{{\rm S}} &  w_{{\rm S}aS} &  w_{{\rm S}aS} & t_{{\rm S}}^* & r_{{\rm S}}^* \\
  w_{{\rm S}} & -w_{{\rm S}} &  w_{{\rm S}aS} &  w_{{\rm S}aS} & r_{{\rm S}}^* & t_{{\rm S}}^* \\
 \end{array}\right)
  \left(\begin{array}{l}
 a_{k} \\ a_{-k} \\ c_{q_{{\rm aS}}} \\ c_{-q_{{\rm aS}}} \\ c_{q_{{\rm S}}}^\dag \\ c_{-q_{{\rm S}}}^\dag
 \end{array}\right)^{\rm in} .
\end{align}

\subsection{Transfer matrices in the basis of right- and left-going waves}

\begin{figure}[t]
  \includegraphics[width=.2\columnwidth]{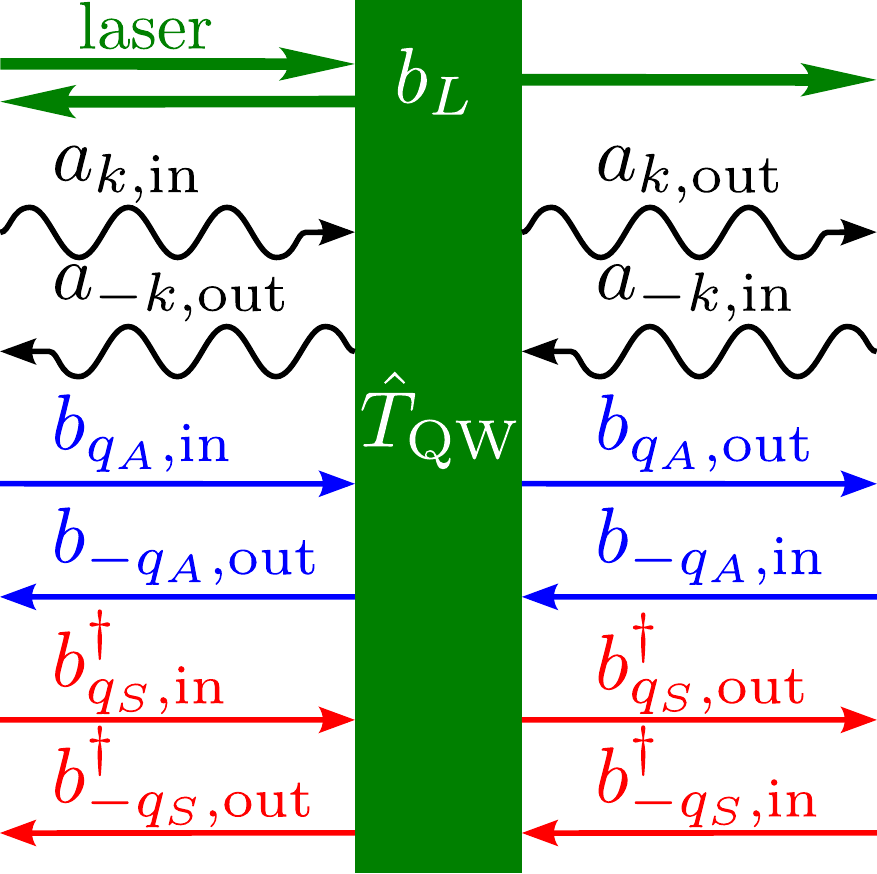}
\caption{An illustration of the transfer matrix relating phonon and photon fields on the right and left edges of the laser-pumped QW.}\label{fig:TQW}
\end{figure}

The transfer matrix relates the phonon and photon operators at the right side of the QW to those at the left side, see Fig.~\ref{fig:TQW},
\begin{align}\label{eq:T}
 \left(\begin{array}{l}
 a_{k}^{\rm out} \\ a_{-k}^{\rm in} \\ c_{q_{{\rm aS}}}^{\rm out} \\ c_{-q_{{\rm aS}}}^{\rm in} \\ c_{q_{{\rm S}}}^{{\rm out}\,\dag} \\ c_{-q_{{\rm S}}}^{{\rm in}\,\dag}
 \end{array}\right) = 
\hat T_{\rm QW}
  \left(\begin{array}{l}
 a_{k}^{\rm in} \\ a_{-k}^{\rm out} \\ c_{q_{{\rm aS}}}^{\rm in} \\ c_{-q_{{\rm aS}}}^{\rm out} \\ c_{q_{{\rm S}}}^{{\rm in}\,\dag} \\ c_{-q_{{\rm S}}}^{{\rm out}\,\dag}
 \end{array}\right).
\end{align}
The transfer matrix for a QW is easily obtained from the scattering matrix~Eq.~\eqref{eq:SQW},
\begin{align}\label{eq:Tw}
  \hat T_{\rm QW}=\left(\begin{array}{rrrrrr}
 1-\frac{2 \rmi \gamma  \Delta }{\Delta ^2-\Omega ^2} & \frac{2 \rmi \gamma  \Delta }{\Delta ^2-\Omega ^2} & -\frac{\e^{-\rmi  \phi } \sqrt{\gamma  \Gamma_0 }}{\Delta +\Omega } & -\frac{\e^{-\rmi  \phi } \sqrt{\gamma  \Gamma_0 }}{\Delta +\Omega } & -\frac{\e^{\rmi \phi } \sqrt{\gamma  \Gamma_0 }}{\Delta -\Omega } & -\frac{\e^{\rmi \phi } \sqrt{\gamma  \Gamma_0 }}{\Delta -\Omega } \\
 -\frac{2 \rmi \gamma  \Delta }{\Delta ^2-\Omega ^2} & 1+\frac{2 \rmi \gamma  \Delta }{\Delta ^2-\Omega ^2} & -\frac{\e^{-\rmi  \phi } \sqrt{\gamma  \Gamma_0 }}{\Delta +\Omega } & -\frac{\e^{-\rmi  \phi } \sqrt{\gamma  \Gamma_0 }}{\Delta +\Omega } & -\frac{\e^{\rmi \phi } \sqrt{\gamma  \Gamma_0 }}{\Delta -\Omega } & -\frac{\e^{\rmi \phi } \sqrt{\gamma  \Gamma_0 }}{\Delta -\Omega } \\
 \frac{\e^{\rmi \phi } \sqrt{\gamma  \Gamma_0 }}{\Delta +\Omega } & -\frac{\e^{\rmi \phi } \sqrt{\gamma  \Gamma_0 }}{\Delta +\Omega } & 1-\frac{\rmi  \Gamma_0 }{\Delta +\Omega } & -\frac{\rmi  \Gamma_0 }{\Delta +\Omega } & 0 & 0 \\
 -\frac{\e^{\rmi \phi } \sqrt{\gamma  \Gamma_0 }}{\Delta +\Omega } & \frac{\e^{\rmi \phi } \sqrt{\gamma  \Gamma_0 }}{\Delta +\Omega } & \frac{\rmi  \Gamma_0 }{\Delta +\Omega } & 1+\frac{\rmi  \Gamma_0 }{\Delta +\Omega } & 0 & 0 \\
 -\frac{\e^{-\rmi  \phi } \sqrt{\gamma  \Gamma_0 }}{\Delta -\Omega } & \frac{\e^{-\rmi  \phi } \sqrt{\gamma  \Gamma_0 }}{\Delta -\Omega } & 0 & 0 & 1+\frac{\rmi  \Gamma_0 }{\Delta -\Omega } & \frac{\rmi  \Gamma_0 }{\Delta -\Omega } \\
 \frac{\e^{-\rmi  \phi } \sqrt{\gamma  \Gamma_0 }}{\Delta -\Omega } & -\frac{\e^{-\rmi  \phi } \sqrt{\gamma  \Gamma_0 }}{\Delta -\Omega } & 0 & 0 & -\frac{\rmi  \Gamma_0 }{\Delta - \Omega} & 1-\frac{\rmi  \Gamma_0 }{\Delta - \Omega} \\
 \end{array}\right).
\end{align}
The transfer matrix for a spacer,  accounting for the phase gained by the light and sound during their passage through the spacer, reads
\begin{align}\label{eq:Tspacer}
  \hat T_{\rm spacer}=\left(\begin{array}{rrrrrr}
 \e^{\rmi kd} & 0 & 0 & 0 & 0 & 0 \\
 0 & \e^{-\rmi kd} & 0 & 0 & 0 & 0 \\
 0 & 0 & \e^{\rmi q_{{\rm aS}}d} & 0 & 0 & 0 \\
 0 & 0 & 0 & \e^{-\rmi q_{{\rm aS}}d} & 0 & 0 \\
 0 & 0 & 0 & 0 & \e^{-\rmi q_{{\rm S}}d} & 0 \\
 0 & 0 & 0 & 0 & 0 & \e^{\rmi q_{{\rm S}}d} \\
 \end{array}\right).
\end{align}
The transfer matrix of the structure $\hat T$ is the product of the transfer matrices of the individual layers, QWs and spacers.
The structure transfer matrix calculated, the left-to-right and right-to-left phonon transmission coefficients $t_\rightarrow$ and $t_\leftarrow$ can be calculated from the matrix equations
\begin{align}\label{eq:Ttrl}
\hat T
  \left(\begin{array}{l}
 1 \\ r_{\hookleftarrow} \\ 0 \\ w_{{\rm S}\hookleftarrow}  \\ 0 \\  w_{{\rm aS}\hookleftarrow} 
 \end{array}\right)
 =
 \left(\begin{array}{l}
 t_\rightarrow \\ 0 \\ w_{{\rm S}\rightarrow} \\ 0 \\ w_{{\rm aS}\rightarrow} \\ 0
 \end{array}\right)
 ; \quad
 \hat T
  \left(\begin{array}{l}
 0 \\ t_{\leftarrow} \\ 0 \\ w_{{\rm S}\leftarrow}  \\ 0 \\  w_{{\rm aS}\leftarrow} 
 \end{array}\right)
 =
 \left(\begin{array}{l}
 r_{\hookrightarrow} \\ 1 \\ w_{{\rm S}\hookrightarrow} \\ 0 \\ w_{{\rm aS}\hookrightarrow} \\ 0
 \end{array}\right).
\end{align}
Expressing $t_\rightarrow$ and $t_\leftarrow$ from these equations we obtain Eq.~\eqref{eq:tt} where $\hat A_{\rightarrow(\leftarrow)}$ denotes the $3\times3$ matrix formed by the elements of matrix $\hat A$ with odd (even) row and columns indices.

The QW transfer matrix has a simpler form in the basis of the fields and their derivatives, $(a,a', c_{{\rm aS}}, c_{{\rm aS}}', c_{\rm S}^\dag, c_{\rm S}'^\dag)$. Here $f = f_{q,\rm in(out)} + f_{-q,\rm out(in)}$ and $f' =\rmi( f_{q,\rm in(out)} - f_{-q,\rm out(in)})$ at the left (right) edge, $f$ is $a$, $c_{{\rm aS}}$, or $c_{\rm S}$. The transfer matrix in such basis, Eq.~\eqref{eq:TQW}, is obtained from that in the basis of traveling waves, Eq.~\eqref{eq:Tw},  by transform $\hat U \hat T_{\rm QW} \hat U^\dag$, where the unitary matrix $\hat U$ reads 
\begin{align}\label{eq:U}
U = 
 \left(\begin{array}{rrrrrr}
  1 & 1 & 0 & 0 & 0 & 0 \\
  \rmi & -\rmi & 0 & 0 & 0 & 0 \\
  0 & 0 & 1 & 1 & 0 & 0 \\
  0 & 0 & \rmi & -\rmi & 0 & 0 \\
  0 & 0 & 0 & 0 & 1 & 1 \\
  0 & 0 & 0 & 0 & -\rmi & \rmi \\
 \end{array}\right) .
\end{align}

\subsection{Derivation of the transmission coefficients for two QWs Eq.~\eqref{eq:dt}}

\begin{figure}[t]
  \includegraphics[width=.6\columnwidth]{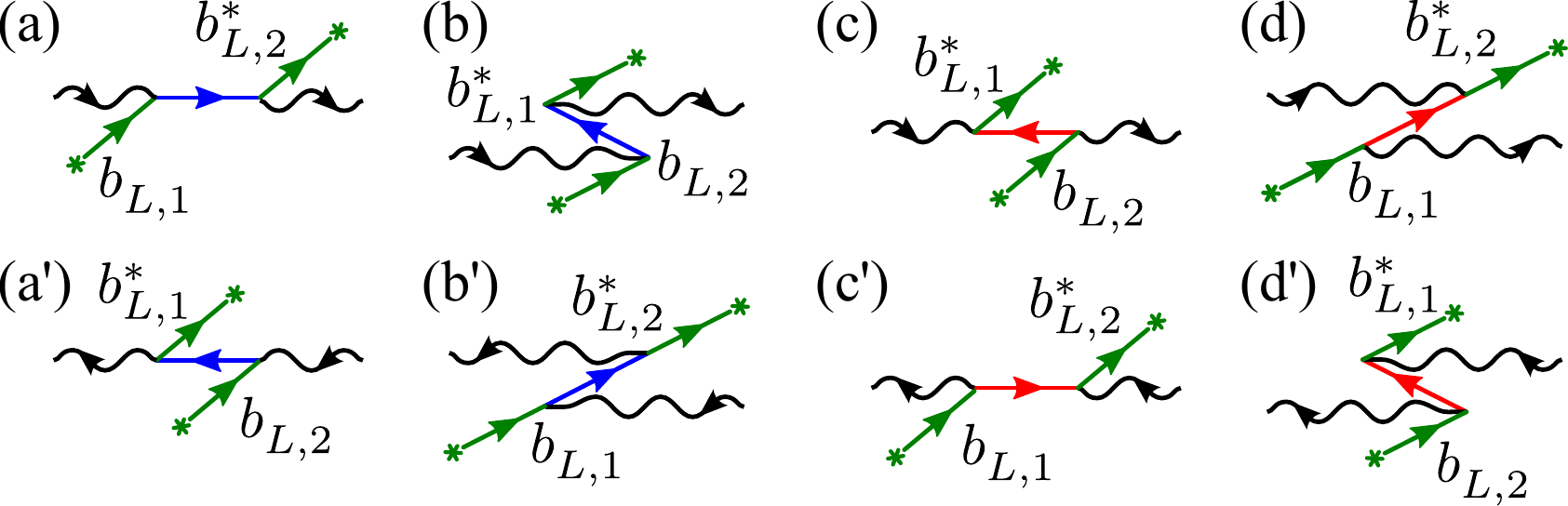}
\caption{(a)-(d) Four elementary processes contributing to the phonon transmission coefficient $t_\rightarrow$ through a pair of QWs. The reciprocal processes contributing to transmission coefficient $t_\leftarrow$ are shown in panels (a$'$)-(d$'$). Wavy lines describe the incident and transmitted phonons, straight  red (blue) lines correspond to the \mbox{(anti-)}Stokes light propagation between the two QWs, an incoming (outgoing) line with a star stands for the laser-generated exciton polarization inside the QW $b_L$ ($b_L^*$), the vertices represent exciton-mediated photoelastic interaction. 
}\label{fig0_s}
\end{figure}

The diagrams showing linear in both $\gamma$ and $\Gamma_0$ contributions to the transmission coefficient $t_\rightarrow$ through a pair of QWs are depicted in Figs.~\ref{fig0_s}(a)-(d). They correspond to a phonon transformed to Stokes or anti-Stokes light in one QW, then propagating to the other QW where it is converted back into the phonon. Summing this amplitudes with the amplitude of transmission without conversion we obtain
\begin{align}
 t_{\rightarrow} &= \e^{\rmi kd} \nonumber
 - \frac{\Gamma_0 \gamma\e^{\rmi(\phi_1-\phi_2)}}{(\Delta+\Omega+\rmi\Gamma)^2} \e^{\rmi q_{{\rm aS}} d} 
 - \frac{\Gamma_0 \gamma\e^{\rmi(\phi_2-\phi_1)}}{(\Delta+\Omega+\rmi\Gamma)^2} \e^{\rmi q_{{\rm aS}} d} \e^{2\rmi kd}
 + \frac{\Gamma_0 \gamma\e^{\rmi(\phi_2-\phi_1)}}{(\Delta-\Omega-\rmi\Gamma)^2} \e^{-\rmi q_{{\rm S}} d} 
 + \frac{\Gamma_0 \gamma\e^{\rmi(\phi_1-\phi_2)}}{(\Delta-\Omega-\rmi\Gamma)^2} \e^{-\rmi q_{{\rm S}} d} \e^{2\rmi kd} \\
  &= \e^{\rmi kd} \left[ 1 - \frac{\gamma\Gamma_0}{(\Delta+\Omega+\rmi\Gamma)^2} \cos(kd+\phi_2-\phi_1)\e^{\rmi q_{{\rm aS}} d} 
  +\frac{\gamma\Gamma_0}{(\Delta-\Omega-\rmi\Gamma)^2} \cos(kd+\phi_1-\phi_2)\e^{-\rmi q_{{\rm S}} d} \right] \,.
  \label{eq:st}
\end{align}
Neglecting $q_{{\rm aS}}d,q_{\rm S} d \ll 1$ we obtain $|t_{\rightarrow}|^2=1+\delta|t_{\rightarrow}|^2$, where $\delta|t_{\rightarrow}|^2$ is given by
Eq.~\eqref{eq:dt} of the main text. The contributions to reciprocal transmission coefficient $t_\leftarrow$ are shown in Fig.~\ref{fig0_s}(a$'$)-(d$'$). The corresponding amplitudes are obtained from Eq.~\eqref{eq:st} by the replacement $\phi_1 \leftrightarrow \phi_2$. The difference of left-to-right and right-to-left transmittances, depicted in Fig.~\ref{fig:tN}(d) is then given by
\begin{align}
\left|\frac{t_\rightarrow}{t_\leftarrow}\right|^2=1+ 16\gamma\Gamma_0 \sin kd \sin(\phi_2-\phi_1) {\rm Re}\frac{(\Omega+\rmi\Gamma)^2+\Delta^2}{[(\Omega+\rmi\Gamma)^2-\Delta^2]^2} \,.
  \end{align}

\subsection{Phonoritons in the infinite structure}

\begin{figure*}[t]
  \includegraphics[width=.99\textwidth]{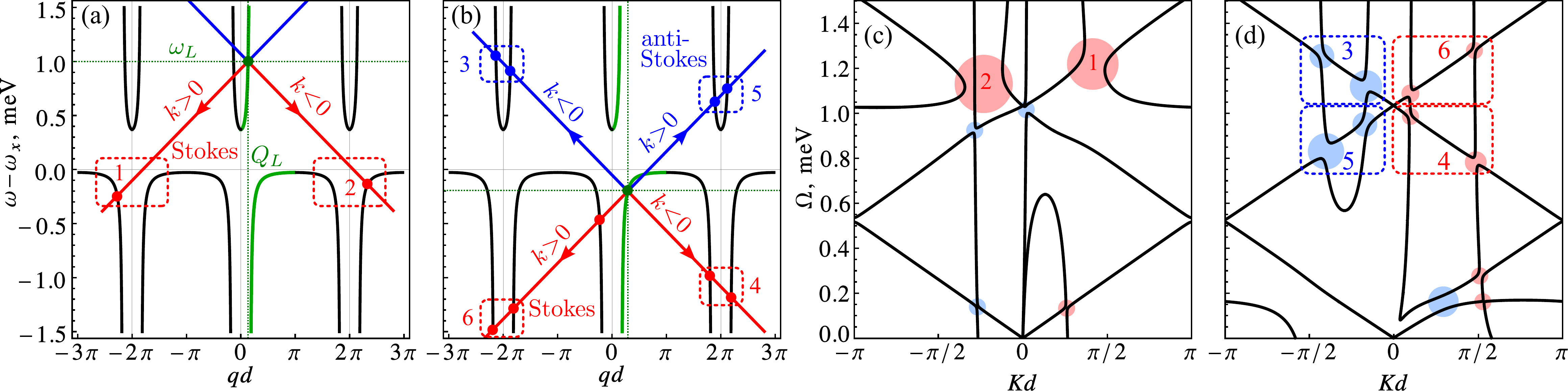}
\caption{(a),(b) Polariton Brillouin scattering processes contributing to the phonon transmission amplitude through a long structure. Black curves show the  polariton dispersion, red and blue lines show the emitted and absorbed phonon dispersion. 
(c),(d) Phonoriton dispersion. Blue and red disks indicate the avoided crossings at the intersection points of the phonon and polariton dispersions.  The numbered areas in panels (a)--(d) explain the corresponding patterns in the phonon transmission map Fig.~\ref{fig:tN}(d). The calculation is made for the same parameters as for the unfolded phonoriton dispersion Fig.~\ref{fig:fig_j}.
}\label{fig:phonor}
\end{figure*}

We calculate here the phonoriton spectrum using the $6\times 6$ transfer matrix technique. To this end we use the transfer matrix through the structure period in the basis where the optical wave phase is measure with respect to the laser-generated polariton phase, $\hat T_1 = \hat T_{\rm QW} \hat T_{\rm spacer} \hat T_{\rm laser}$. Here the latter matrix accounts for the laser phase and reads
\begin{align}\label{eq:Tlaser}
  \hat T_{\rm laser}=\left(\begin{array}{cccccc}
 \hphantom{\e}1\hphantom{\e} & 0 & 0 & 0 & 0 & 0 \\
 0 & \hphantom{\e}1\hphantom{\e} & 0 & 0 & 0 & 0 \\
 0 & 0 & \e^{-\rmi Q_L d} & 0 & 0 & 0 \\
 0 & 0 & 0 & \e^{-\rmi Q_L d} & 0 & 0 \\
 0 & 0 & 0 & 0 & \e^{\rmi Q_L d} & 0 \\
 0 & 0 & 0 & 0 & 0 & \e^{\rmi Q_L d} \\
 \end{array}\right).
\end{align}
The transfer matrix through $N$ periods in the considered basis is simply $\hat T_N = \hat T_1^N$, while the dispersion of the eigen phonoriton modes $K(\Omega)$ is found from  
\begin{align}
 {\rm det}[\hat T_1(\Omega) - \e^{\rmi K d} \hat I] = 0 \,,
\end{align}
 where $\hat I$ is the identity matrix. The obtained with this procedure phonoriton dispersion is shown in Fig.~\ref{fig:phonor}(c) and~(d).  Graphically, phonoriton dispersion Fig.~\ref{fig:phonor}(c)-(d) can be obtained by superposing the corresponding scattering plot Figs.~\ref{fig:phonor}(a)-(b) and its copy inverted with respect to the laser point $(Q_L, \omega_L)$, marked by the green dot dot, followed by the reduction to the first Brillouin zone. The superposition and folding leads to several crossings that are in fact avoided crossings of two types, as discussed in the main text. Figure~\ref{fig:fig_j}(a)-(b) of the main text is obtained by unfolding the phonoriton dispersion Fig.~\ref{fig:phonor}(c)-(d).  

Far from the Brillouin zone edges, the phonoriton dispersion is well described by the self energy correction Eq.~\eqref{eq:phondisp}.
In the vicinity of a crossing of the polariton dispersion with the dispersion of absorbed (emitted) phonon, the phonoriton dispersion equation reads
\begin{align}
(\Omega-\Omega_k)\left(\omega_L \pm \Omega-\omega_x-\Gamma_0\frac{\sin q_{\rm aS(S)}d}{\cos(Q_L\pm k)d-\cos q_{\rm aS(S)}d} \right) = \gamma \frac sd  \,.
\end{align}
Linearizing polariton dispersion in the vicinity of the crossing we obtain the splitting value Eq.~\eqref{eq:delta}.

\end{document}